\documentclass[smallextended]{svjour3}       % onecolumn (second format)
\smartqed  % flush right qed marks, e.g. at end of proof

\usepackage{graphicx}
\usepackage{amssymb}
\usepackage{mathrsfs, amsmath, float, multirow}
\usepackage[ruled, vlined, lined, linesnumbered]{algorithm2e}
\usepackage{graphicx, tabularx, subfigure, subfloat}
\usepackage{url}
\begin{document}
\newcommand*{\QEDA}{\hfill\ensuremath{\blacksquare}}  % for proof
\newcommand*{\QEDB}{\hfill\ensuremath{\square}}  % for proof
\newcommand\T{\rule{0pt}{2.6ex}}
\newcommand\B{\rule[-1.2ex]{0pt}{0pt}}

\title{Survival analysis, the infinite Gaussian mixture model, FDG-PET and non-imaging data in the prediction of progression from mild cognitive impairment%\thanks{Grants or other notes
}

\titlerunning{GMM and model selection for PET classification}        % if too long for running head

\author{Rui Li
\and Robert Perneczky
\and Alexander Drzezga
\and Stefan Kramer
\and for the Alzheimer's Disease Neuroimaging Initiative
}


\institute{R. Li (co-first author) \at
              Institut f\"ur Informatik/I12, Technische Universit\"at M\"unchen, Boltzmannstr. 3, 85748 Garching b. M\"unchen, Germany \\
              \email{rui.li@in.tum.de}           %  \\
            \and
            R. Perneczky (co-first author) \at
            Neuroepidemiology and Ageing Research Unit, School of Public Health, Faculty of Medicine,
            The Imperial College of Science, Technology and Medicine, St Dunstan's Road, London, W6 8RP, UK\\
            Klinik und Poliklinik f\"ur Psychiatrie und Psychotherapie, Technische Universit\"at M\"unchen, Ismaninger Str. 22, 81675 M\"unchen, Germany\\
            West London Cognitive Disorders Treatment and Research Unit, West London Mental Health Trust, London, Brentford Lodge, Boston Manor Road, London, TW8 8DS, UK\\
            \email{r.perneczky@imperial.ac.uk}
            \and A. Drzezga \at
            Klinik und Poliklinik f\"ur Nuklearmedizin, Universit\"at zu K\"oln, Kerpener Str. 62, 50937 K\"oln, Germany\\
            \email{alexander.drzezga@uk-koeln.de}
           \and
           S. Kramer (corresponding author) \at
              Institut f\"ur Informatik, Johannes Gutenberg - Universit\"at Mainz, Staudingerweg 9,  55128 Mainz, Germany\\
              Tel.: +49-6131-39-23378\\
              Fax: +49-6131-39-23534\\
              \email{kramer@informatik.uni-mainz.de}
}


\maketitle

\begin{abstract}
We present a method to discover interesting brain regions in [18F] fluorodeoxyglucose positron emission tomography (PET) scans, showing also the benefits when PET scans are in combined use with non-imaging variables. The discriminative brain regions facilitate a better understanding of Alzheimer's disease (AD) progression, and they can also be used for predicting conversion from mild cognitive impairment (MCI) to AD. A survival analysis (Cox regression) and infinite Gaussian mixture model (IGMM) are introduced to identify the informative brain regions, which can be further used to make a prediction of conversion (in two years) from MCI to AD using only the baseline PET scan. Further, the predictive accuracy can be enhanced when non-imaging variables are used together with identified informative brain voxels. The results suggest that PET scan imaging data is more predictive than other non-imaging data, revealing even better performance when both imaging and non-imaging data are combined.
\keywords{Survival analysis \and Cox regression \and MCI conversion \and PET \and infinite Gaussian mixture model \and Clustering}
\end{abstract}

\section{Introduction}
\label{intro}

Alzheimer's disease (AD) is a progressive, degenerative and incurable disease of the brain and the most prevalent cause of dementia. The number of people suffering from dementia is expected to grow rapidly in the next decades due to increasing life expectancy \cite{ALZ08}, which will have a major negative impact on health care systems worldwide. Despite technological progress, the \emph{ante mortem} diagnosis of AD is still based on clinical grounds, with biomarkers such as cerebrospinal fluid (CSF) proteins and neuroimaging procedures providing supporting information.

Positron emission tomography with [18F] fluorodeoxyglucose (FDG-PET) has been widely applied to assist in the diagnosis of clinical AD \cite{Drzezga09}. Accordingly, FDG-PET was recommended as a diagnostic marker by recently proposed guidelines \cite{Dubois07}. One important area in which functional neuroimaging may add significant value is the prediction progression of prodromal AD, i.e., mild cognitive impairment (MCI), to full-blown AD dementia. A growing number of studies show that FDG-PET can predict the clinical outcome in MCI with a relatively high sensitivity and specificity. A recent literature review underpins the gain in overall diagnostic accuracy by using FDG-PET in the evaluation of dementia, supporting its role as an effective complementary tool \cite{Bohnen12}. Superiority of FDG-PET to other potential predictors of clinical decline in MCI has been suggested by some \cite{Landau10}, but not all \cite{Gomar14} studies. Typically, an AD-like FDG-PET pattern with regional decreases in cerebral glucose metabolism in posterior temporoperietal regions and the posterior cingulate cortex, as well as to a lesser degree the prefrontal cortex, is found to be predictive of future AD dementia in patients with MCI \cite{Mosconi08}.

Despite the compelling evidence in favor of FDG-PET as a prognostic marker, most experts concur that there is an immediate need for further efforts to improve implementation of neuroimaging in current diagnostic paradigms, including the optimization of the image analysis methods. Previous studies suggest that improved analytical methods such as principal component analysis \cite{Habeck08}, linear programming discriminant analysis \cite{Yu14}, support vector machines (SVM), Gaussian process classification \cite{Young13} or tree structured sparse learning \cite{Liu14} may improve the overall diagnostic and prognostic classification of patients. The main aim of the present study is of two-fold: first, to apply survival analysis and the infinite Gaussian mixture model to FDG-PET scans in order to improve the differentiation between patients who remain in the MCI stage ($\text{MCI}_\text{MCI}$) and those that progress to AD dementia ($\text{MCI}_\text{AD}$), and second, to investigate the usefulness of non-imaging data compared to FDG-PET imaging data.

\section{Materials and Methods}
\subsection{Study Participants}

Experiments were performed using data of the publicly available AD Neuroimaging Initiative (ADNI) database (http://adni.loni.usc.edu/) accessed in the year 2011. Only data from the first stage of ADNI (ADNI 1) were considered. ADNI has a large pool of FDG-PET (co-registered, averaged) images, which have been acquired on various scanners using different imaging parameters. To eliminate bias due to these factors, we selected images that had been obtained using the same scanner (Siemens/CTI) as well as the same parameters, such as the number of slices. To ensure that differences in the degree of cognitive impairment at baseline did not affect our results, we only included patients with a Clinical Dementia Rating (CDR) sum of the boxes score of 1.5, 2 and 2.5 \cite{Morris93}. Patients were only considered if they had at least five consecutive clinical assessments every six months up to a maximum of 24 months in order to have sufficient data for the survival analysis. The planned pre-selection of participants resulted in a total of 77 patients, including 45 $\text{MCI}_\text{MCI}$ and 32 $\text{MCI}_\text{AD}$. $\text{MCI}_\text{MCI}$ was defined as patients not meeting the Institute of Neurological and Communicative Disorders and Stroke-AD and Related Disorders Association (NINCDS-ADRDA) criteria for AD at their last follow-up assessment, whereas $\text{MCI}_\text{AD}$ patients met the NINCDS-ADRDA criteria at least at one of the follow-up visits. It is important to note that this is a \emph{censored} dataset, which includes patients that did not progress to AD dementia until their last follow-up visit  for various reasons including withdrawal of consent, death or limited length of follow-up. This phenomenon is often studied using \emph{survival analysis} \cite{Kleinbaum11}, which is also the essential technique applied in this work. The characteristics of the study population are summarized in Table 1.

\begin{table*}
\centering
\caption{Baseline information mean (SD) of the patients. The protocol of ADNI diagnostic and image acquisition is described in appendix.CDR: Clinical Dementia Rating. CDT: clock drawing test, 5 is the best score and 0 is the worst. MMSE: Mini-Mental State Examination. ADAS: Alzheimer's Disease Assessment Scale. The higher the ADAS, the more severe of mental illness. More explanation is referred to http://www.adni-info.org/scientists/Pdfs/ADNI\_GeneralProceduresManual.pdf}{
\label{table:baselineInformation}
\begin{tabular}{lcccccc}
\hline\T
 & Subjects  & Age & CDR Total & CDT & MMSE & ADAS\\
  & (female:male) &  &   &  &  & \\
  \hline\T
  $\text{MCI}_\text{MCI}$ & 45 (13:32) & 76 (8.2) & 1.97 (0.3) & 4.3 (0.8) & 27.4 (1.6) & 15 (6.6)\\
    $\text{MCI}_\text{AD}$ & 32 (12:20) & 75 (6.9) & 1.95 (0.3) & 3.8 (1.1) & 26.7 (1.7) & 19 (4.6)\\
  \hline
\end{tabular}}
\end{table*}

Time to progression from MCI to AD dementia was calculated as the time between the baseline visit and the visit at which an AD dementia diagnosis was first established. As a result, we obtain a dataset that allows us to train a model predicting progression to AD dementia within 24 months.

\subsection{Image Pre-processing}
Prior to their use for image analysis, the FDG-PET images underwent two pre-processing steps in the statistical parametric mapping software package SPM5 (Wellcome Functional Imaging Laboratory, London, UK), based on Matlab R2010a (The Mathworks Inc, Natick, USA): spatial normalization and smoothing (kernel size [8 8 8] mm). Spatial normalization ensures that the processed image is of the size 91$\times$109$\times$91 voxels, which is in accordance with the Anatomical Automatic Labeling (AAL) template \cite{Tzourio-Mazoyer02}. The final step is intensity normalization, which was done by dividing each voxel by the mean intensity value averaged over the primary sensorimotor cortex region, which has been shown to improve FDG-PET based discrimination compared to other brain regions or the global metabolic mean \cite{Yakushev08}. Anatomically, the ``Precentral\_L, Precentral\_R, Postcentral\_L and Postcentral\_R" regions in the AAL brain template can be used as the primary sensorimotor cortex. The analysis is performed only on the AAL defined brain regions (gray matter voxels of MNI space). In total, there are 185,405 voxels in the AAL brain template. Among 185,405 voxels in each FDG-PET scan, there is a portion of voxels that contain discriminative information. Hence, the following method is proposed to discover the discriminative voxels.

\subsection{Identification of discriminative Voxels}

Cox regression is suitable to be applied when some censored records occur in a dataset, which is exactly the scenario of the ADNI follow-up study (end of study after, for example, 5 visits). Therefore, we use this technique in a first step to select the discriminative voxels of FDG-PET scans.

\subsubsection{Introduction to Cox Regression (Survival Analysis)}

Cox regression is a semi-parametric survival analysis method. It makes no assumption about the probability distribution of the survival time, assuming only proportional hazards. Cox regression is also known as the Cox proportional hazards model \cite{Cox72}. The rate at which the failure happens or the patient suffers from a disease is known as the hazard function. Let $x_1$, $x_2$,..., $x_p$ be the values of $p$ covariates $X_1$, $X_2$,..., $X_p$. The hazard function is defined as:

\begin{equation}\label{eq:hazardfunction}
    h(t) = h_0(t)\exp\left(\sum_{i=1}^p \beta_i x_i \right),
\end{equation}

where $\beta_1$,  $\beta_2$,..., $\beta_p$ is the $1 \times p$ vector of regression parameters and $h_0(t)$ is the baseline hazard function at that time. The coefficient vectors of the covariates are estimated using a maximum likelihood (ML) estimate, which is obtained by maximizing a partial likelihood function. The hazard function focuses on failing, whereas the survivor function focuses on ``surviving" given survival up to a certain time point. The hazard function $h(t)$ and survivor function   $s(t)$ can be derived from each other. The general formula for their relation is:

\begin{equation}\label{eq:survivalfunction}
    s(t) = \exp\left(-\int_0^t h(u)du\right).
\end{equation}

Since we apply Cox regression on each single voxel in this first step, we have only one covariate $\beta_1$ to determine, and $x_1$ is the voxel intensity. Kleinbaum \cite{Kleinbaum11} offers a more comprehensive introduction to survival analysis.

\subsubsection{Cox Regression applied to FDG-PET}

Given the baseline PET scan, we run the Cox regression on each of the 185,405 voxels (AAL defined) independently on 77 studied samples, collecting the ones that show a negative correlation $\beta_1$ with hazardness at a $p$-value smaller than 0.01 (i.e., voxels that correlate positively with the survival time). The statistical significance level of 0.01 instead of 0.05 is chosen because we apply no multiple comparisons correction. The reason why we did not perform multiple comparisons correction is that too many voxels may be discarded after correction. The resulting voxels are first filtered out in this way, before they are combined in a classification model (see below).

\subsubsection{Selection of neighboring Voxels by Infinite Gaussian Mixture Model}
A number of discriminative (informative) voxels are identified by the Cox regression analysis. Since an informative voxel's neighboring voxels also tend to be informative (in part due to the partial volume effect \cite{Rousset98}, we need to eliminate such an effect to avoid overfitting. We therefore applied the infinite Gaussian mixture model to divide the voxels into clusters based on their $x$, $y$, and $z$ coordinates (the resulting clusters are illustrated in Fig. 1).

The infinite Gaussian Mixture Model (IGMM) \cite{Rasmussen00} was proposed as an extension of the widely applied Gaussian Mixture Model (GMM) \cite{Bishop06}. In a GMM, the number of clusters (components) is assumed to be fixed a priori, which is, in fact, hard to do in practice. On the contrary, IGMM assumes the number of clusters is unknown (infinity, not to limit the number), which is determined by the data in the end. The IGMM can be briefly written as follows:

\begin{figure}
  \includegraphics[width=1\textwidth]{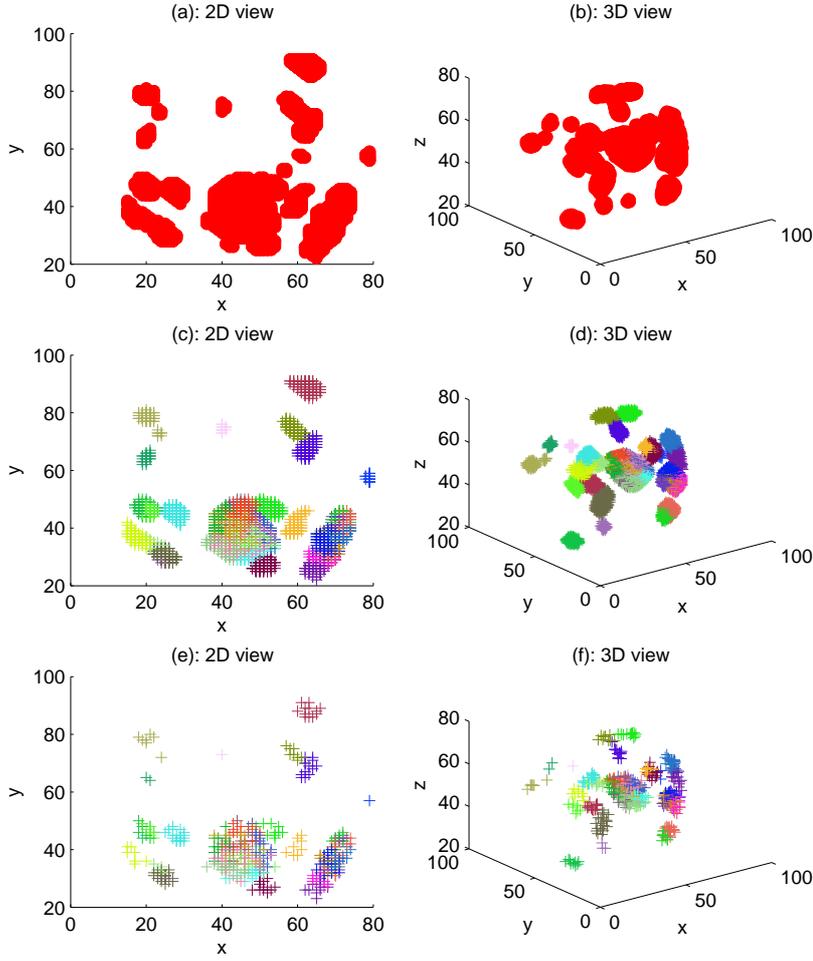}
\caption{Illustration of significant voxels and the clustering results. (a): 3D view of the same voxels shown in 2D. (b): 3D view of significant voxels after Cox regression. (c): 2D view of clustered voxels (same color represents same cluster) after applying the infinite Gaussian mixture model. (d): 3D view of the same voxels shown in subplot (c). (e): 2D view of selected 10\% voxels shown in subplot (c) and (d). (f): 3D view of the same voxels shown in subplot (e).}
\label{fig:Fig1}
\end{figure}

\begin{figure}
  \includegraphics[width=1\textwidth]{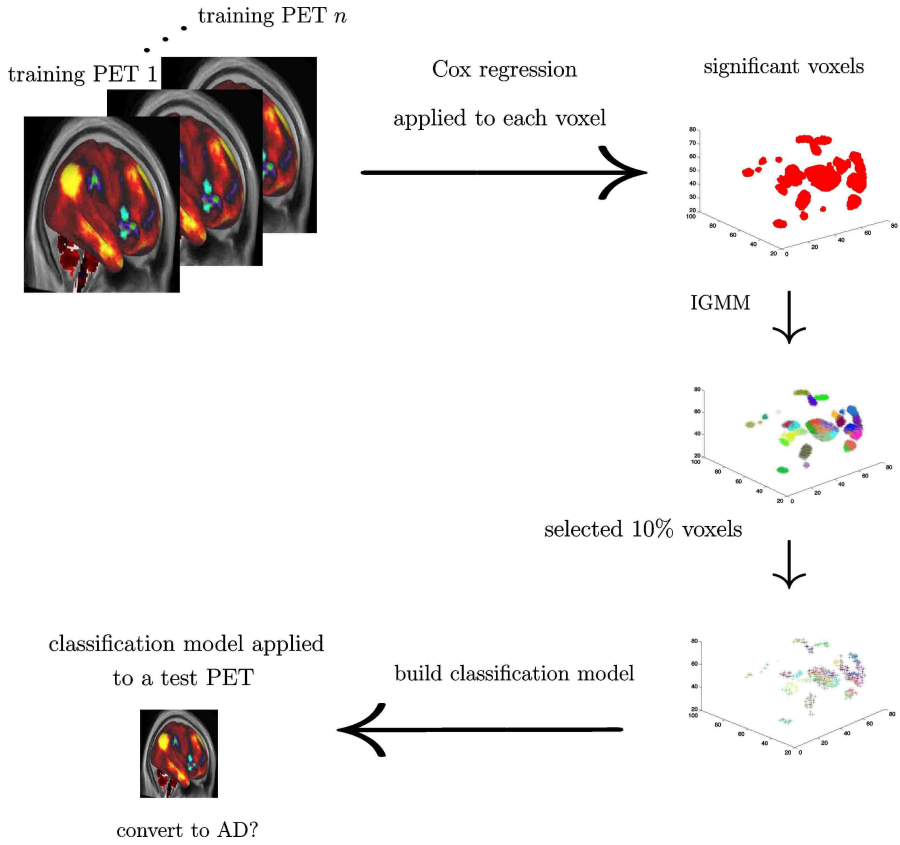}
\caption{Workflow of proposed MCI conversion prediction. IGMM: infinite Gaussian mixture model.}
\label{fig:Fig2}
\end{figure}

\begin{equation}\label{eq:IGMM1}
    p(x|\theta) = \sum_{k=1}^{K}\pi_k \mathcal{N}(x;\mu_k, \mathrm{\Sigma}_k),
\end{equation}

where $\mu_k$, $\mathrm{\Sigma}_k$ and $\pi_k$ are the mean, covariance and mixing proportion respectively. In addition, $\mathrm{\Sigma}_{k=1}^K \pi_k = 1$, $\pi_k\ge 0$ and $\theta = \{\mu_k, \mathrm{\Sigma}_k, \pi_k\}$. $\mathcal{N}$ denotes the $D$-dimensional Gaussian distribution:

\begin{equation}\label{eq:IGMM2}
   \mathcal{N}(X|\mu, \mathrm{\Sigma}) = \frac{1}{(2\pi)^{\frac{D}{2}}|\mathrm{\Sigma}|^\frac{1}{2}}\exp\left(-\frac{1}{2}(X-\mu)^T
   \mathrm{\Sigma}^{-1}(X-\mu) \right).
\end{equation}

By allowing $K\rightarrow \infty$, IGMM extends the GMM in terms of the number of clusters. The inference is achieved by Markov chain Monte Carlo (MCMC), performing Gibbs sampling for a number of iterations. In the end, the voxel data neighbouring each other are likely to be grouped in one cluster.

After applying IGMM, the clustered voxels in one cluster are adjacent to each other. We subsequently chose 10\% of the voxels in each cluster to represent the respective cluster, selecting voxels spread widely across the cluster. Finally, we collected the chosen 10\% of voxels in every cluster as the discriminative voxels for the prediction model. In our experiments, 10\% is empirically shown to be optimal. The reason may be that, too few voxels, such as 1\%, may exclude some discriminative ones hence causing underfitting. On the other hand, too many voxels, such as 50\% may still cause overfitting. Another reason is that the complementary information conveyed by other non-imaging data (see Section ``Building the Classification Model") cannot be fully used, due to curse of dimensionality, if too many voxel features are selected in building the classification model. The workflow is depicted in Fig. 2.

\subsection{Building the Classification Model}
The identified imaging voxels were used as features to build a classification model. In this study, we employed support vector machines (SVM), which is a state-of-the-art classifier. LIBSVM \cite{Chang11} was used to build the SVM models with a linear kernel with grid search for parameter optimization. Grid search considers only the optimization of the penalty parameter in the linear SVM, selecting the value that yields the best classification result based on the training data. After the best value is found, it is applied to the test data.
To validate the model, the whole dataset was split into two subsets, a training set used for model building and a test set used to test the performance of the model. A 5-fold cross-validation was applied to split the data, which was achieved by dividing it into five disjoint subsets, with four subsets as training and the remaining subset as test dataset. In addition to the imaging data (FDG-PET), we also investigated the non-imaging data (cf. Table 1) with the aim to consider information derived from more than just one source or modality. The non-imaging data includes age, gender, the results of the Clock Drawing Test (CDT), the Mini-Mental State Examination (MMSE) and the Alzheimer's Disease Assessment Scale (ADAS). Note that we did not use the Clinical Dementia Rating (CDR) in building the model, because CDR is a stronger indicator used for the diagnosis.

\section{Results}
The study samples show that the number of males is greater than the number of females in both $\text{MCI}_\text{MCI}$ and $\text{MCI}_\text{AD}$ (cf. Table 1). The two groups of $\text{MCI}_\text{MCI}$ and $\text{MCI}_\text{AD}$ indicate similar average age and CDR at baseline. Other cognitive indicators, such as CDT, MMSE, ADAS, all suggest that the $\text{MCI}_\text{MCI}$ group is at a better cognitive status than $\text{MCI}_\text{AD}$.

\subsection{Brain Voxels identified by Cox Regression Analysis}

\begin{figure}
  \includegraphics[width=1\textwidth]{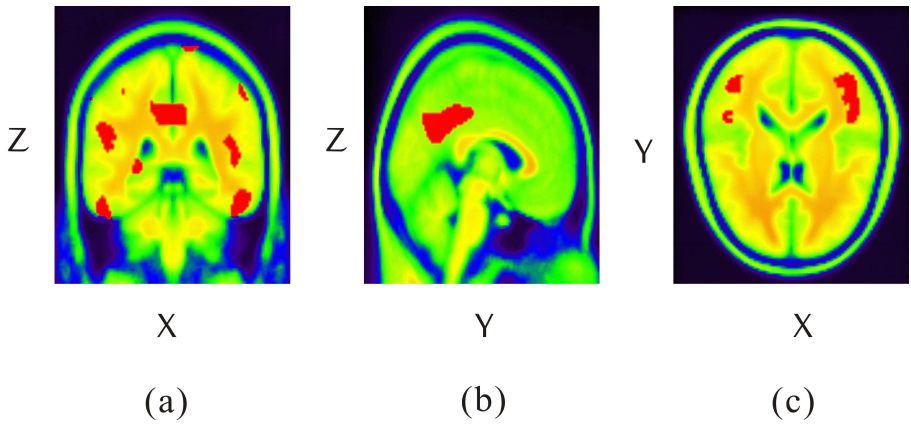}
\caption{2D view of the interesting regions at the 45th layer based on the whole dataset using Cox regression at a $p$-value of 0.01 (a): coronal view (b): sagittal view (c): transaxial view. $x$, $y$ and $z$ are the width (91), depth (109) and height (91) respectively. The red points represent the informative voxels (brain regions).}
\label{fig:Fig3}
\end{figure}

\begin{table*}
\centering
\caption{The top-10 interesting regions and their proportions using the whole dataset. The proportion is calculated as the number of significant voxels in the region divided by the number of total significant voxels. Region name is referred to AAL template \cite{Tzourio-Mazoyer02}.}{
\label{table:top10region}
\begin{tabular}{lclc}
\hline\T
  Regions & Percentage & Regions & Percentage \\
  \hline\T
  1: Precuneus\_L & 13.9\% & 6: Temporal\_Mid\_L & 5.79\% \\
  2: Precuneus\_R &11.3\% & 7: Parietal\_Sup\_R & 5.05\%\\
  3: Parietal\_Inf\_L & 10.8\% & 8: Parietal\_Inf\_R & 4.71\%\\
  4: Angular\_R & 10.3\% & 9: Parietal\_Sup\_L & 3.84\%\\
  5: Angular\_L & 7.78\% & 10: Cingulum\_Mid\_R & 3.57\%\\
  \hline
\end{tabular}}
\end{table*}

After applying a Cox regression analysis, a number of discriminative voxels served as essential features to train the subsequent model. Some identified voxels are displayed in Fig. 3. Table 2 reveals these voxels' associated regions defined in the AAL template. The ``Precuneus\_L, Precuneus\_R, Parietal\_Inf\_L and Angular\_R" account for particularly high percentages compared to other regions. It is known that the precuneus is involved in several essential cognitive tasks. For example, episodic memory, visual-spatial abilities, and motor activity coordination strategies. The parietal lobe includes symbolic functions in language and numbers and interpretation of spatial information. The remaining identified regions, such as angular, temporal and cingulum, are associated with some cognitive abilities as language, mathematics and memory, etc.

\subsection{Prediction of Progression to AD Dementia}

\begin{figure}
\centering
  \includegraphics[width=0.8\textwidth]{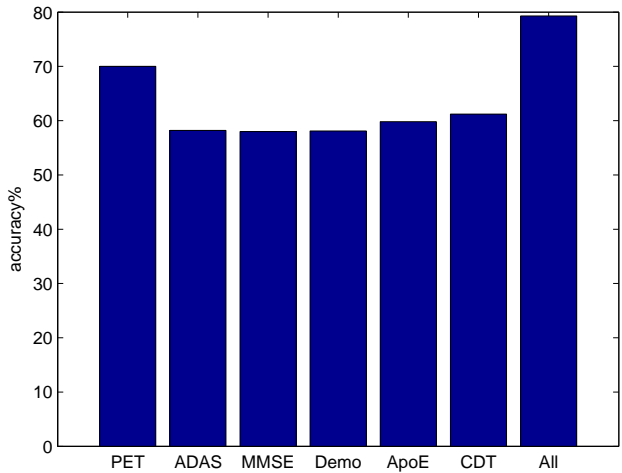}
\caption{Comparison of classification accuracy of various information sources. All: using all data (PET, ADAS, MMSE, Demo, ApoE and CDT) as features in SVM. ApoE: Apolipoprotein E.}
\label{fig:Fig4}
\end{figure}

The baseline accuracy of random guessing is at 45/(45+32) = 58.4\% (45 $\text{MCI}_\text{MCI}$ and 32 $\text{MCI}_\text{AD}$) to predict that a patient does not progress to AD dementia within two years. Fig. 4 demonstrates that the FDG-PET scan alone achieves a classification accuracy of 70\%, which is higher than the accuracy of any other information source. Classification accuracy reaches nearly 80\% when all the sources of information are pooled together to build the predictive model, the reason being that the model benefits from gaining complementary discriminative information from diverse sources.

\section{Discussion}
Accurately predicting the course of MCI is an important clinical and academic aim, but censored data often limit the ability to derive meaningful results. Survival analysis is a suitable statistical tool for this kind of situation since it allows to analyze incomplete datasets. The current study corroborates findings of previous studies, showing that a metabolic deficit in the temporoparietal cortex, the precuneus and the limbic cortex offer valuable information in terms of dementia risk in MCI \cite{Mosconi08}. These brain areas are well known to play a role in episodic memory, visuospatial processing and executive function, which all are typically affected early in the course of AD \cite{Weintraub12}. Our results also show that survival analysis is a viable statistical method to discriminate between progressive and stable MCI cases.
The usual limitations of studies based on clinical cohorts recruited at specialized memory clinics apply to our study. These include the lack of generalizability of the results to the wider population due to highly selected MCI patients with a high a \emph{priori} probability of suffering from AD; the lack of histopathological verification of the clinical diagnoses; the limited follow-up period; and the restricted sample size.
We also demonstrate the benefits of a clustering algorithm (IGMM) to cluster the significant voxels derived from the Cox regression model. These voxels can be clustered into different groups with respect to their geometric similarity. By choosing a portion (i.e., 10\%) of the voxels in each cluster, we avoid using too many voxels and thereby effectively reduce the risk of overfitting, while still maintaining informative voxels for dementia prediction. We also corroborate the view that combining imaging data with neurocognitive and demographical information leads to improved classification accuracy. To combine various sources of data, one may also use multi-view stacking \cite{Li11}. However, the present research does not benefit from it because of the limited classification performance of non-imaging variables. In our case, collecting all data into a simple form for learning yields satisfactory results. The improvement in accuracy can be attributed to the complementary information provided by non-imaging variables, in other words, more complementary information sources together benefit the training model.

\section{Conclusions}
The present study proposes a survival analysis method to analyze neuroimaging data  to gain insights into the ability of FDG-PET to predict conversion from MCI to AD dementia within 24 months. By treating data from patients not progressing to AD dementia as censored records, we were able to use survival analysis to detect the brain regions that convey discriminative information. This approach revealed brain regions that are typically associated with early clinical AD and can be used to predict progression in MCI. The use of IGMM divides the discriminative voxels into various clusters, coping with overfitting by selecting a portion of voxels. Less discriminative power is conveyed by the neurocognitive and demographical data, which however still seem to provide complementary information, which altogether improves the prediction model. To conclude, our study describes the application of robust statistical methods to FDG-PET data analysis that may improve the use of neuroimaging data for the characterization of MCI in prospective research settings.

\begin{flushleft}
\textbf{Information Sharing Statement}
\end{flushleft}

Data used in preparation of this article were obtained from the Alzheimer's Disease Neuroimaging Initiative (ADNI) database (adni.loni.usc.edu). As such, the investigators within the ADNI contributed to the design and implementation of ADNI and/or provided data but did not participate in analysis or writing of this report. A complete listing of ADNI investigators can be found at:
\url{http://adni.loni.usc.edu/wp-content/uploads/how_to_apply/ADNI_Acknowledgement_List.pdf}

\begin{acknowledgements}
Rui Li acknowledges the support of the TUM Graduate School of Information Science in Health (GSISH), Technische Universit\"at M\"unchen.

Data collection and sharing for this project was funded by the Alzheimer's Disease Neuroimaging Initiative (ADNI) (National Institutes of Health Grant U01 AG024904) and DOD ADNI (Department of Defense award number W81XWH-12-2-0012). ADNI is funded by the National Institute on Aging, the National Institute of Biomedical Imaging and Bioengineering, and through generous contributions from the following: Alzheimer's Association; Alzheimer's Drug Discovery Foundation; BioClinica, Inc.; Biogen Idec Inc.; Bristol-Myers Squibb Company; Eisai Inc.; Elan Pharmaceuticals, Inc.; Eli Lilly and Company; F. Hoffmann-La Roche Ltd and its affiliated company Genentech, Inc.; GE Healthcare; Innogenetics, N.V.; IXICO Ltd.; Janssen Alzheimer Immunotherapy Research \& Development, LLC.; Johnson \& Johnson Pharmaceutical Research
\& Development LLC.; Medpace, Inc.; Merck \& Co., Inc.; Meso Scale Diagnostics, LLC.; NeuroRx
Research; Novartis Pharmaceuticals Corporation; Pfizer Inc.; Piramal Imaging; Servier; Synarc Inc.; and Takeda Pharmaceutical Company. The Canadian Institutes of Health Research is providing funds to support ADNI clinical sites in Canada. Private sector contributions are facilitated by the Foundation for the National Institutes of Health (www.fnih.org). The grantee organization is the Northern California Institute for Research and Education, and the study is coordinated by the Alzheimer's Disease Cooperative Study at the University of California, San Diego. ADNI data are disseminated by the Laboratory for NeuroImaging at the University of Southern California.
\end{acknowledgements}

\begin{flushleft}
\textbf{Conflict of Interest} The authors have no conflicts of interest.
\end{flushleft}

\begin{flushleft}
\textbf{Appendix: ADNI Diagnostic and Image Acquisition Protocol}
\end{flushleft}

The ADNI recruitment and inclusion procedures are described in detail at www.adni-info.org. Briefly, at baseline, subjects in ADNI were between 55$-$90 years of age, had a modified Hachinski score $\le$ 4 and at least six years of education. Patients with MCI had MMSE scores between 24 and 30, a CDR score of 0.5; they had memory complaints, but no significant functional impairment, and objective memory deficits on the Wechsler Memory-Scale-Logical Memory II test. After the baseline visit, follow-up visits were conducted at six- or 12-month intervals up to a maximum of six years. FDG-PET were acquired within two weeks before or two weeks after the in-clinic assessments at Baseline and at the second annual visit, 24 months after Baseline.


%\renewcommand{\BBAA}{\&} % edited by Rui Li
\bibliographystyle{plain}
\bibliography{LPDK}

\end{document}